\documentclass
[twocolumn,aps,prl,amsmath,amssymb,floatfix,superscriptaddress,nofootinbib,preprintnumbers]{revtex4-2}%
\usepackage[utf8]{inputenc}
\usepackage{graphicx}
\usepackage{bm}
\usepackage{mathrsfs}
\usepackage{mathtools}
\usepackage{color}
\usepackage[section]{placeins}
\usepackage{hyperref}
\usepackage{amsmath}
\usepackage{amsfonts}
\usepackage{amssymb}%
\hypersetup{hidelinks}

\begin{document}
\preprint{CTP-SCU/2026002}
\title{Early-Time Nonlinear Growth in an Unstable Q-Ball Hairy Black Hole}
\author{Lang Cheng}
\email{langcheng@stu.scu.edu.cn}
\affiliation{College of Physics, Sichuan University, Chengdu 610065, China}
\author{Guangzhou Guo}
\email{guangzhou@sjtu.edu.cn}
\affiliation{Tsung-Dao Lee Institute, Shanghai Jiao Tong University, Shanghai 201210, China}
\author{Peng Wang}
\email{pengw@scu.edu.cn}
\affiliation{College of Physics, Sichuan University, Chengdu 610065, China}
\author{Haitang Yang}
\email{hyanga@scu.edu.cn}
\affiliation{College of Physics, Sichuan University, Chengdu 610065, China}

\begin{abstract}
Early-time evolution away from an unstable equilibrium in a nonlinear system
is often expected to be governed by the associated linear instability.
Combining full nonlinear evolution with first- and second-order quasinormal
mode (QNM) calculations, we show that this expectation can fail during the
unstable growth stage of a Q-ball hairy black hole in Einstein-Maxwell theory
with a charged self-interacting scalar field. The linear unstable QNM has a
much larger amplitude in one component of the scalar field than in the other:
the more strongly responding component follows that mode, whereas the early
growth of the more weakly responding component is dominated by a second-order
QNM sourced by the linear unstable mode. This occurs while the evolution
remains perturbative. Our results thus show that the early growth of an
individual component need not be governed by its linear response.

\end{abstract}
\maketitle

\noindent\textbf{\textit{Introduction.}} Linear perturbation theory often
provides the natural starting point for describing an unstable system as it
departs from equilibrium. In a nonlinear system with several coupled dynamical
components, however, \textquotedblleft early time\textquotedblright%
\footnote{Here and throughout, ``early time'' refers to the perturbative
unstable growth stage before nonlinear saturation, rather than to the earliest
coordinate times after initialization.} need not imply linear dominance in
every component. If a linear instability excites different components with
strongly unequal amplitudes, then a quadratic response sourced by the more
strongly excited component can overtake the linear growth of the more weakly
responding one while the evolution still remains within the perturbative
regime. In that case, the early prominent growth observed in a given component
need not follow its linear response. Establishing such a scenario requires a
framework in which both the nonlinear evolution and the relevant perturbative
modes can be computed explicitly.

Black hole (BH) dynamics provide such a framework, since they are accessible
at both the fully nonlinear and perturbative levels. Nonlinear evolutions of
BH spacetimes, especially hairy BHs, have revealed rich dynamics beyond linear
theory
\cite{Sanchis-Gual:2015lje,Sanchis-Gual:2016tcm,Sanchis-Gual:2016ros,Herdeiro:2018wub,Doneva:2021dcc,Zhang:2021nnn,Zhang:2022cmu,Liu:2022eri,Chen:2022vag,Chen:2023eru,Chen:2023iws,Jiang:2023yyn,Garcia-Saenz:2024beb,Zhang:2025jlb,Melis:2024kfr,Guo:2024cts,Garcia-Saenz:2025rbc,Qin:2026axh}%
, while perturbative analyses can resolve not only the linear quasinormal
modes (QNMs) but also additional nonlinear contributions generated by mode
couplings
\cite{Campanelli:1998jv,Nakano:2007cj,Berti:2009kk,Kehagias:2023ctr,Khera:2023oyf,Ma:2024qcv,Bucciotti:2024zyp,Bourg:2024jme,Bucciotti:2024jrv,Khera:2024bjs,Pan:2024bon,Bourg:2025lpd,Berti:2025hly}%
. Recent waveform studies further show that subdominant nonlinear
contributions can be isolated from time-domain signals
\cite{Ma:2022wpv,Cheung:2022rbm,Mitman:2022qdl,Cheung:2023vki,Giesler:2024hcr,Mitman:2025hgy,Yang:2025ror,Wang:2026rev}%
. This leads to a sharper question than whether nonlinear effects are merely
present: during the unstable growth of a multicomponent BH system, does the
growth of a given component really track its linear response, or is it already
governed by a nonlinear contribution?

In this Letter, we address this question in the nonlinear evolution of an
unstable Q-ball hairy BH. In Einstein-Maxwell theory coupled to a charged
scalar field with a polynomial self-interaction, static BHs with hair
supported by the complex scalar field exist in addition to the bald
Reissner-Nordstr\"{o}m (RN) branch \cite{Hong:2020miv,Herdeiro:2020xmb}. These
solutions do not bifurcate continuously from RN BHs, but instead exhibit a
finite gap and form two branches that meet at a turning point
\cite{Herdeiro:2020xmb}. Subsequent nonlinear evolutions further suggested
that the two branches have distinct dynamical properties, with one being
unstable under small perturbations and the other serving as a dynamically
robust end state \cite{Zhang:2023qtn}. This makes the system well suited for
examining how linear and quadratic responses compete between the two
components of the complex scalar field during the unstable growth stage of a
fully dynamical BH evolution.

Here we show that, during the unstable growth stage of a Q-ball hairy BH, the
evolution of a given component need not track its linear response. For the
complex scalar deviation from equilibrium, the imaginary part is dominated by
the linear unstable QNM. The real part, by contrast, is the more weakly
responding component and, although it contains a linear contribution, is
governed through most of the growth stage by a quadratic response sourced by
the linear unstable mode. Our results thus provide a clear example, in a fully
dynamical BH system, of how a weakly responding component can already be
dominated at early times by a quadratic response.

\begin{figure*}[t]
\centering
\includegraphics[width=0.95\textwidth]{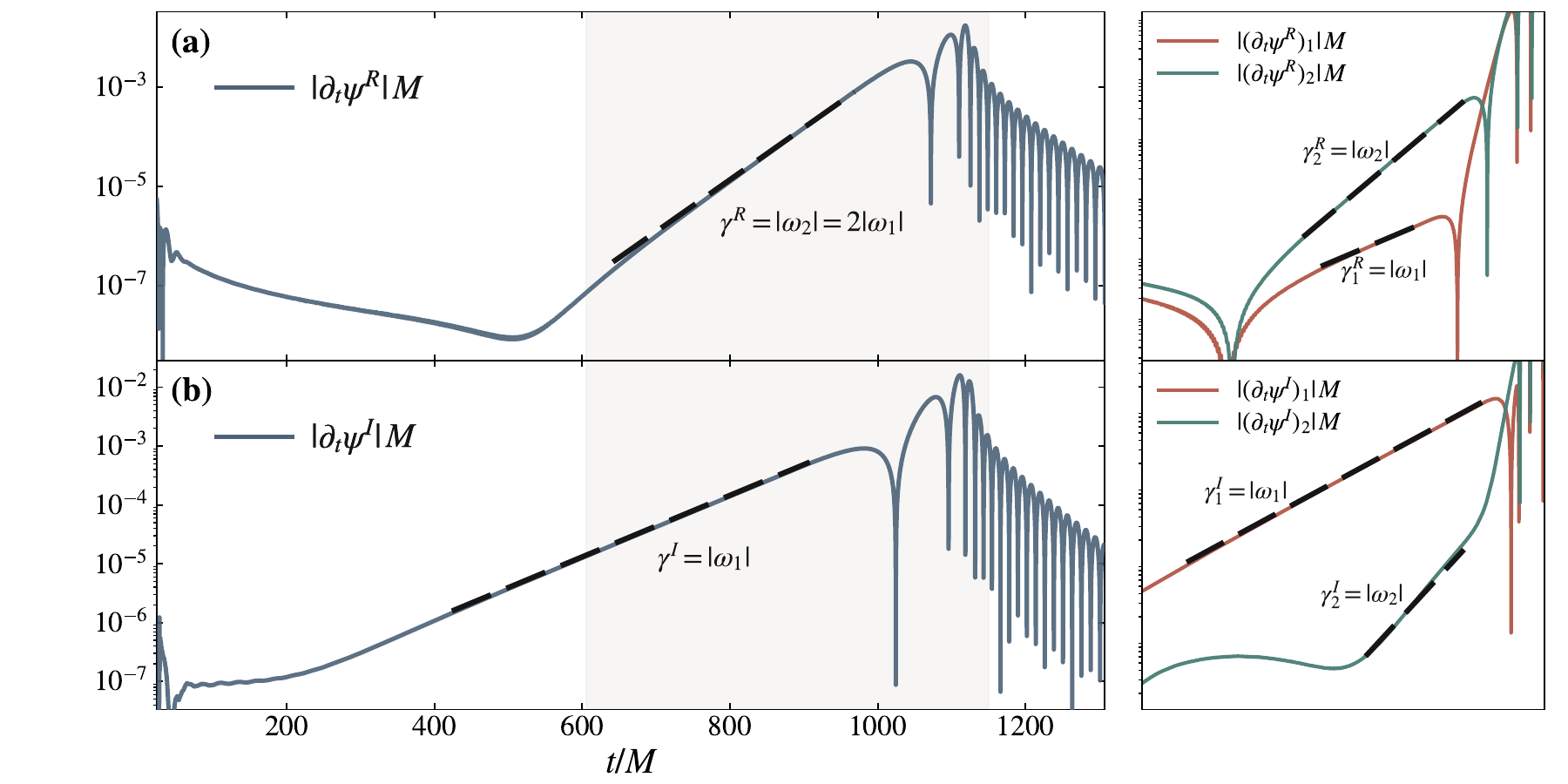}\caption{Time-domain
evolution at the BH horizon for an unstable Q-ball hairy BH. The left panels
show $\left\vert \partial_{t}\psi^{R}\right\vert $ and $\left\vert
\partial_{t}\psi^{I}\right\vert $, where $\psi^{R}$ and $\psi^{I}$ denote the
real and imaginary parts of the scalar deviation from the initial equilibrium
profile. The shaded regions indicate the time windows used for the QNM filter.
The black dashed lines indicate the growth rates associated with the linear
unstable mode and the quadratic contribution, with frequencies $\omega_{1}$
and $\omega_{2}$, respectively. The right panels display the magnitudes of the
filtered quantities: the upper panel shows $\left\vert (\partial_{t}\psi
^{R})_{1}\right\vert $ and $\left\vert (\partial_{t}\psi^{R})_{2}\right\vert
$, while the lower panel shows $\left\vert (\partial_{t}\psi^{I}%
)_{1}\right\vert $ and $\left\vert (\partial_{t}\psi^{I})_{2}\right\vert $.
These quantities isolate the contributions from $\omega_{1}$ and $\omega_{2}$,
respectively. The filtered profiles indicate that the imaginary part is
controlled by the linear unstable mode, whereas the growth of the real part is
dominated by the quadratic contribution. The linear contribution in the real
part becomes visible only after filtering.}%
\label{fig:filter}%
\end{figure*}

\noindent\textbf{\textit{Model and nonlinear evolution.}} We consider
Einstein-Maxwell gravity coupled to a charged, self-interacting scalar field,
\begin{equation}
S=\frac{1}{16\pi}\int\mathrm{d}^{4}x\sqrt{-g}\,\left[  R-F^{2}-2|D\Psi
|^{2}-U(|\Psi|^{2})\right]  , \label{eq:action-main}%
\end{equation}
with $F^{2}\equiv F_{ab}F^{ab}$ and $|D\Psi|^{2}\equiv(D_{a}\Psi)^{\ast}%
D^{a}\Psi$, where $D_{a}=\nabla_{a}-iqA_{a}$. The scalar self-interaction
potential is taken to be
\begin{equation}
U(|\Psi|^{2})=m^{2}|\Psi|^{2}+\beta|\Psi|^{4}+\gamma|\Psi|^{6}.
\label{eq:potential-main}%
\end{equation}
Here, $q$ determines the scalar charge, while $m$, $\beta$, and $\gamma$
specify the polynomial self-interaction potential. Throughout this Letter, we
use geometrized units with $G=c=4\pi\epsilon_{0}=1$. In the absence of a BH,
the self-interacting scalar sector admits Q-balls, namely nontopological
solitons supported by the complex scalar field. They arise naturally in
self-interacting scalar theories, have long been discussed as dark matter
candidates, and have been widely studied in perturbative, nonlinear, and
superradiant settings
\cite{Friedberg:1976me,Coleman:1985ki,Lee:1991ax,Kusenko:1997si,Enqvist:1997si,Kusenko:1997vp,Axenides:1999hs,Battye:2000qj,Kusenko:2001vu,Smolyakov:2017axd,Smolyakov:2019cld,Saffin:2022tub,Cardoso:2023dtm,Ciurla:2024ksm,Chen:2025oxo}%
.

Here we focus on spherically symmetric BH solutions of the form
\begin{equation}
\begin{aligned} \mathrm{d}s^{2}&=-\left[ 1-2m_{t}\left( r\right) /r\right] \mathrm{d}t^{2}+\frac{\mathrm{d}r^{2}}{1-2m_{r}\left( r\right) /r}+r^{2}\mathrm{d}\Omega_{2}^{2},\\ A&=V(r)\mathrm{d} t,\qquad \Psi=\Psi_0(r). \end{aligned} \label{eq:bg-main}%
\end{equation}
For this class of static charged BHs, one can use the residual $U(1)$ gauge
freedom, together with the regular horizon gauge condition $V(r_{h})=0$,
to choose $\Psi_{0}(r)$ to be real \cite{Hong:2020miv,Herdeiro:2020xmb}.
Details of the construction of these spherically symmetric hairy BH solutions
are given in the Supplementary Material. The Q-ball hairy BH considered in
this paper is specified by $(q,m,\beta,\gamma)=(1.25,1.64,-301.88,9428.84)$
and $(Q,M)=(0.97,1)$, and belongs to the dynamically unstable sector of the
hairy BH family.

We investigate the nonlinear evolution of the unstable hairy BH introduced
above by evolving the full Einstein-Maxwell-scalar system in spherical
symmetry. The evolution is initialized from the corresponding unstable
equilibrium solution and is driven entirely by the intrinsic instability of
that state, with small numerical noise from discretization providing the initial perturbation. No external perturbation is imposed by hand. Our primary interest is the
unstable growth stage, during which the system is driven away from equilibrium
before nonlinear saturation sets in. The metric sector is evolved using the
CCZ3 method \cite{Garcia-Saenz:2025dsr}. The Maxwell and scalar sectors are
formulated in a $3+1$ decomposition and integrated using a finite difference
method of lines scheme, with evolution equations that are first order in time.
Details of the evolved variables and equations are summarized in the
Supplementary Material.

The resulting time-domain evolution is shown in Fig.~\ref{fig:filter}. To
monitor the development of the instability, we decompose the dynamical scalar
field as
\begin{equation}
\Psi(t,r)=\Psi_{0}(r)+\psi^{R}(t,r)+i\,\psi^{I}(t,r). \label{eq:pert-main}%
\end{equation}
Because the equilibrium scalar profile $\Psi_{0}(r)$ is real, $\psi^{R}$ and
$\psi^{I}$ describe deviations parallel and orthogonal to this profile in the
complex scalar plane, respectively. The left panels of Fig.~\ref{fig:filter}
show the time evolution of $|\partial_{t}\psi^{R}|$ and $|\partial_{t}\psi
^{I}|$ at the horizon. The two components show distinct growth patterns. The
imaginary part remains small before entering a visible exponential growth
regime around $t\sim200M$. The real part also stays small initially, but
begins to grow only much later, around $t\sim500M$. Moreover, the
corresponding growth rates appear to differ between $|\partial_{t}\psi^{R}|$
and $|\partial_{t}\psi^{I}|$. At later times, both components leave the growth
regime and enter an oscillatory relaxation toward saturation.

To interpret this behavior, we compute the relevant unstable QNMs
independently in the frequency domain for the same hairy BH that serves as the
initial state of the nonlinear evolution. With the convention $e^{-i\omega t}%
$, a QNM with a purely imaginary frequency and positive imaginary part gives
rise to exponential growth, $e^{-i\omega t}=e^{|\omega|t}$. At first order, we
find a single unstable QNM with frequency $\omega_{1}=0.012\,i$ by solving the
fully coupled linearized perturbation equations for the metric, Maxwell field,
and scalar field. We then compute the second-order mode from the corresponding
sourced perturbation equations, in which the first-order solution enters as a
nonlinear source, and find its frequency to be $\omega_{2}=2\omega_{1}$. These
results provide independent predictions for the growth rates associated with
the first- and second-order modes seen in the nonlinear evolution. Technical
details of the QNM calculation are given in the Supplementary Material.

The black dashed lines in the upper-left and lower-left panels of
Fig.~\ref{fig:filter} indicate the first- and second-order growth rates,
$|\omega_{1}|$ and $|\omega_{2}|$, respectively. During the growth stage, the
instantaneous growth rates $\gamma_{R,I}\equiv d\ln|\partial_{t}\psi
^{R,I}|/dt$ closely follow $\gamma_{I}\sim|\omega_{1}|$ and $\gamma_{R}%
\sim|\omega_{2}|$. This provides direct time-domain evidence that the
imaginary part is controlled by the linear unstable QNM, whereas the real part
is governed primarily by the quadratic response. At the same time, the
evolution remains within the perturbative regime through most of this stage.
At the horizon, for example, we find $\Psi_{0}=0.124423$, while at $t=900M$
the corresponding scalar deviations are $\psi^{R}=-0.00613$ and $\psi
^{I}=-0.03902$. Both remain small compared with the equilibrium profile,
although $\psi^{R}$ is significantly smaller than $\psi^{I}$. Consequently,
the quadratic response can already dominate over the linear growth of
$\psi^{R}$ while the evolution is still perturbative.

To disentangle the linear and quadratic contributions to the horizon signal
more clearly, we apply a QNM filter to the signal within the shaded time
windows indicated in the two left panels of Fig.~\ref{fig:filter}, following
the method developed in Ref.~\cite{Ma:2022wpv}. Starting from the horizon
time-domain signals $\partial_{t}\psi^{R,I}(t,r_{h})$, we construct filtered
signals containing the $\omega_{1}$ and $\omega_{2}$ contributions, denoted by
$(\partial_{t}\psi^{R,I})_{1}$ and $(\partial_{t}\psi^{R,I})_{2}$,
respectively. The filtering procedure is described in the Supplementary
Material. The upper-right and lower-right panels of Fig.~\ref{fig:filter} show
the resulting $\left\vert (\partial_{t}\psi^{R})_{1}\right\vert $ and
$\left\vert (\partial_{t}\psi^{R})_{2}\right\vert $, and $\left\vert
(\partial_{t}\psi^{I})_{1}\right\vert $ and $\left\vert (\partial_{t}\psi
^{I})_{2}\right\vert $, respectively. The filtered quantities $\left\vert
(\partial_{t}\psi^{R,I})_{1}\right\vert $ grow with rate $|\omega_{1}|$, while
$\left\vert (\partial_{t}\psi^{R,I})_{2}\right\vert $ grow with rate
$|\omega_{2}|$, supporting the interpretation that the former are dominated by
the linear unstable mode and the latter by the quadratic response. The
filtered results are therefore consistent with the conclusion that, within the
filtered windows, the imaginary part is dominated by the linear unstable QNM,
whereas in the real part the quadratic contribution is much larger than the
linear one through most of the growth stage.

\begin{figure}[t]
\centering
\includegraphics[width=\columnwidth]{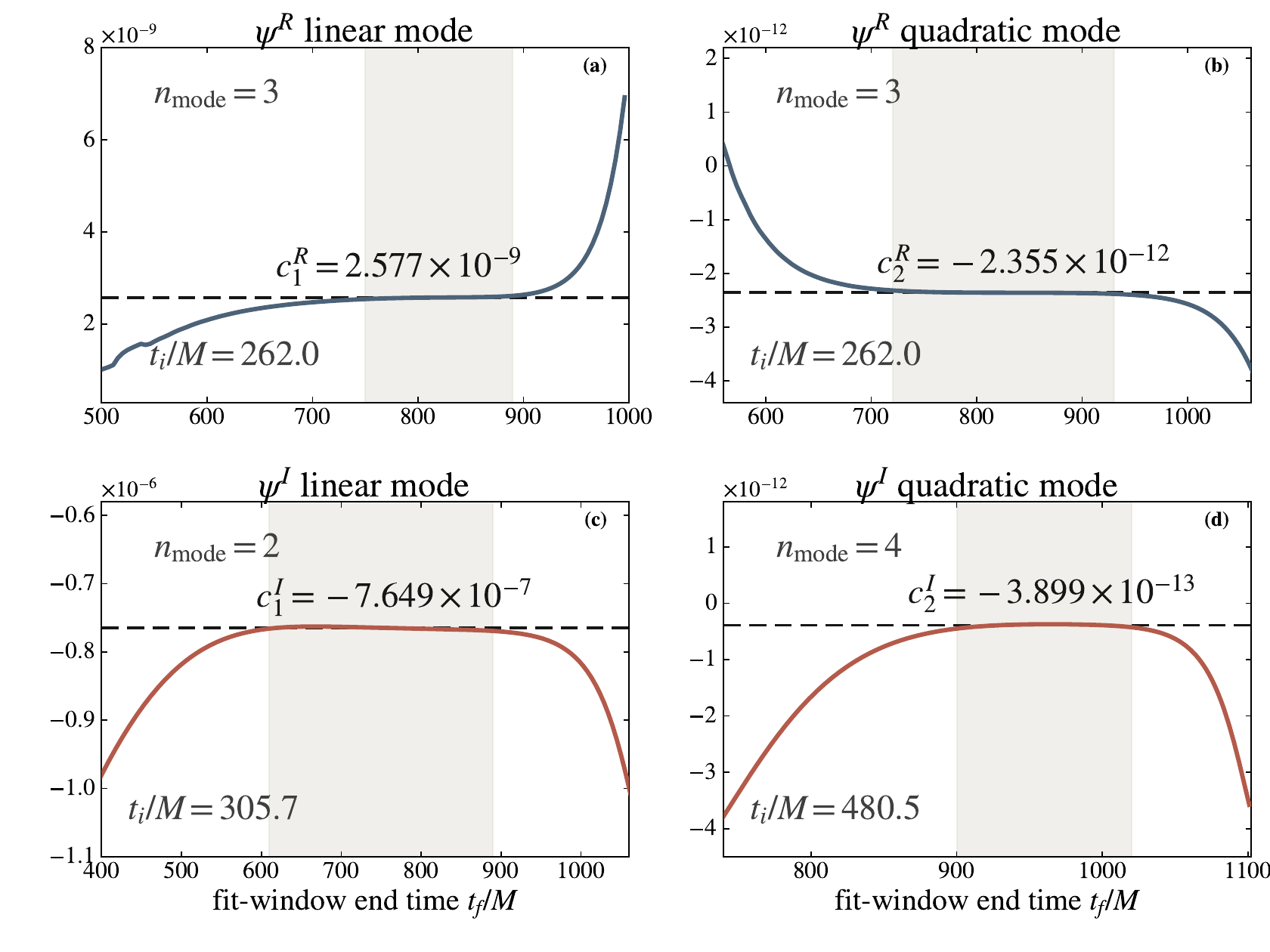}\caption{Coefficient
extraction from QNM fits to the horizon signal $\psi^{R,I}(t,r_{h})$. Panels
(a) and (c) show the extracted linear coefficients $c_{1}^{R}$ and $c_{1}^{I}%
$, while panels (b) and (d) show the extracted quadratic coefficients
$c_{2}^{R}$ and $c_{2}^{I}$, plotted as functions of the fit-window end time
$t_{f}$. In each panel, the start time $t_{i}$ and the maximal QNM order
$n_{\mathrm{mode}}$ are chosen separately. The shaded intervals indicate the
adopted fit windows. The quoted values in each panel are obtained by averaging
over the corresponding shaded interval. The clear, nearly flat plateaus
indicate that the coefficient extraction is robust. The extracted coefficients
further show that the linear contribution is much larger in the imaginary part
than in the real part, whereas the quadratic contribution is smaller in the
imaginary part than in the real part.}%
\label{fig:plateau}%
\end{figure}

\textbf{\textit{QNM extraction and quadratic contribution.}} To quantify the
QNM content of the horizon signal during the unstable growth stage, we fit
$\psi^{R,I}(t,r_{h})$ with
\begin{equation}
\psi^{R,I}(t,r_{h})\sim\sum_{n=1}^{n_{\mathrm{mode}}}c_{n}^{R,I}%
e^{-i\omega_{n}t}, \label{eq:timefit}%
\end{equation}
where $n_{\mathrm{mode}}$ specifies the highest QNM order retained in the fit,
and $\omega_{n}=n\omega_{1}$ is the frequency of the $n$th-order QNM. Since the fitting ansatz is linear in the coefficients $c_{n}^{R,I}$, these coefficients can be determined by a linear least-squares fit. The resulting coefficients quantify the contribution of each QNM order to the real and imaginary parts of the scalar signal at the horizon. To assess the
robustness of the extraction, we repeat the fit over a sequence of time
windows with fixed start time $t_{i}$ and varying end time $t_{f}$. If the
waveform is well described by the QNM expansion, the fitted coefficients
exhibit a stable plateau and remain nearly constant over a range of end times.

Fig.~\ref{fig:plateau} shows the linear and quadratic coefficients extracted
from multi-mode fits to the horizon signals $\psi^{R,I}(t,r_{h})$, plotted as
functions of the fit-window end time $t_{f}$, with the start time $t_{i}$ and
the maximal retained order $n_{\mathrm{mode}}$ chosen separately for each
panel. In all four cases, the fitted coefficients exhibit clear, nearly flat
plateaus over the shaded intervals, indicating that the extraction is robust
within these windows. The quoted coefficients in each panel are obtained by
averaging over the corresponding shaded interval. From these results, we find
that the linear contribution is about two orders of magnitude smaller in the
real part than in the imaginary part. This shows that, at the linear level,
$\psi^{R}$ is the more weakly responding component, whereas $\psi^{I}$ carries
the dominant linear unstable response. By contrast, the quadratic contribution
is about one order of magnitude larger in the real part than in the imaginary
part. This further indicates that the real part, as the more weakly responding
component, can already be dominated at an early stage by the quadratic response.

To compare the linear and quadratic contributions within the same component,
one must take into account not only the extracted coefficients but also their
associated exponential factors. For the imaginary part $\psi^{I}$, the
quadratic contribution is still only about $10\%$ of the linear one at
$t\sim1000M$, indicating that the growth of $\psi^{I}$ is well described by
the linear contribution throughout the preceding stage. The situation is
markedly different for the real part $\psi^{R}$. Around $t\sim500M$, when
$\psi^{R}$ begins to grow, the quadratic contribution has already reached
about $40\%$ of the linear one. By $t\sim700M$, it is already about four times
larger. This shows that the growth of $\psi^{R}$ is much better described by
the quadratic contribution than by the linear one.

\begin{figure}[t]
\centering
\includegraphics[width=\columnwidth]{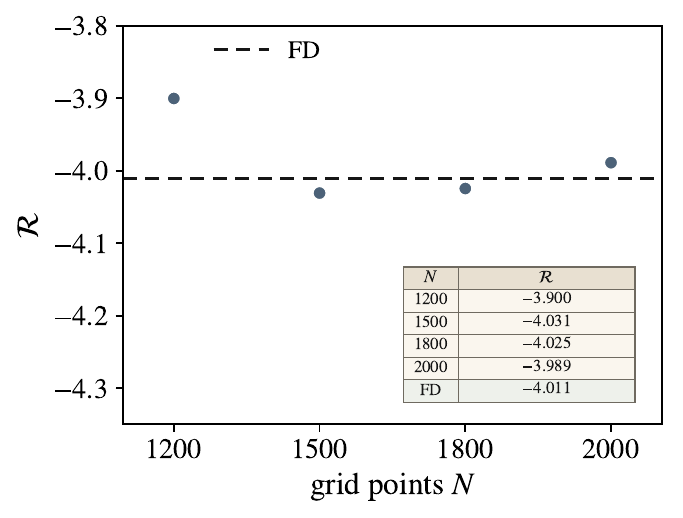}\caption{Dependence
of the ratio $\mathcal{R}\equiv c_{2}^{R}/(c_{1}^{I})^{2}$ on the radial grid
resolution. The points represent the plateau averages extracted from the
selected fitting windows in the time-domain evolutions. The dashed horizontal
line indicates the corresponding frequency-domain prediction, which is
consistent with the values extracted from the time-domain evolutions. The
numerical values of $\mathcal{R}$ are listed in the inset table.}%
\label{fig:ratio}%
\end{figure}

To cross-check the time-domain result, we consider the ratio
\begin{equation}
\mathcal{R}\equiv\frac{c_{2}^{R}}{(c_{1}^{I})^{2}},
\end{equation}
and compare the value extracted from the time-domain evolution with its
frequency-domain counterpart. For the frequency-domain calculation of
$\mathcal{R}$, we first compute the first-order eigenfunctions of the scalar,
gravitational, and Maxwell fields associated with the unstable QNM. Using
these first-order solutions, we then construct the second-order source and
solve the corresponding sourced equations to obtain the second-order
eigenfunctions of the relevant fields. The frequency-domain value of
$\mathcal{R}$ is then obtained by evaluating the second-order eigenfunction of
$\psi^{R}$ and the first-order eigenfunction of $\psi^{I}$ at the horizon, and
taking their ratio as defined above. Further details of the calculation are
given in the Supplementary Material.

Fig.~\ref{fig:ratio} shows the ratio $\mathcal{R}$ extracted from nonlinear
evolutions performed with four different radial grid resolutions, together
with the corresponding frequency-domain prediction. The plotted points
represent the plateau averages obtained from the selected fitting windows,
while the dashed horizontal line indicates the frequency-domain value. In
these evolutions, the instability is triggered solely by intrinsic numerical
noise, so varying the grid resolution effectively changes the size of the seed
perturbation that initiates the unstable growth. Despite this change in the
effective initial perturbation, the extracted values of $\mathcal{R}$ remain
nearly unchanged across the different resolutions. This is precisely what one
expects if the dominant quadratic response scales as $c_{2}^{R}\propto
(c_{1}^{I})^{2}$, namely, if it is sourced by the linear unstable mode. Across
the resolutions considered, the extracted ratios remain close to the
frequency-domain value. This agreement supports the consistency between the
time-domain extraction and the frequency-domain calculation. Taken together,
these results provide a nontrivial cross-check that the prominent early growth
of the more weakly responding real part is a sourced quadratic response.

\noindent\textbf{\textit{Discussion.}} In this Letter, we studied the
nonlinear evolution of an unstable Q-ball hairy BH in Einstein-Maxwell theory
with a charged self-interacting scalar field, combining full time-domain
evolution with first- and second-order frequency-domain calculations. Our
results show that the unstable behavior of the real and imaginary parts of the
scalar field is controlled differently. This distinction arises from a strong
imbalance in how the unstable mode projects onto the two components. The
linear unstable QNM carries a much larger amplitude in the imaginary part than
in the real part, while the corresponding second-order source produces a
comparatively larger response in the real part. As a result, the quadratic
contribution overtakes the linear one in the real part through most of the
growth stage. This shows that \textquotedblleft early time\textquotedblright%
\ and \textquotedblleft linear dominance\textquotedblright\ need not be
identified component by component in a coupled nonlinear system. What fails is
not perturbation theory itself, but the simpler expectation that the early
growth seen in every component should be governed by the linear contribution
in that component.

Within BH physics, this result is closely connected to second-order
perturbation theory and to recent studies of nonlinear ringdown, where
quadratic QNM contributions sourced by dominant linear modes can become
appreciable and, in some channels, even compete with the linear contribution
\cite{Mitman:2022qdl,Ma:2024qcv,Bucciotti:2024zyp}. Our analysis belongs to
this broader line of work, but the emphasis here is different in two respects.
First, the relevant regime is an unstable growth phase rather than a decaying
post-merger ringdown. Second, the competition highlighted here takes place
within a given component of the scalar field, rather than within a radiative
multipole of the emitted signal. Beyond BH physics, related quadratic driving
of weakly responding or otherwise subdominant degrees of freedom, including
mean flows, zonal flows, and low-mode harmonics, is also familiar in
pattern-forming, fluid, and plasma systems
\cite{Rogers2000zonal,Chen2004faraday,Diamond2005zonal,Connaughton2010rossby,Krebs2013lowN,Wang2022zonal}%
.

At the same time, the scope of the present result should be stated carefully.
Our analysis is performed for a specific unstable Q-ball hairy BH in spherical
symmetry, and the main diagnostics are extracted from the horizon signal
during the unstable stage. The present work therefore provides a clean and
fully computable example, rather than a general statement about multicomponent
unstable systems. Whether the same behavior occurs elsewhere depends on the
detailed linear projection of the unstable mode and on the structure of the
second-order source. It would therefore be valuable to examine whether
analogous competition between weak linear and stronger sourced quadratic
responses also arises in less symmetric settings, particularly in rotating
backgrounds and in observables defined at null infinity.

\FloatBarrier
\bigskip\noindent\textbf{Acknowledgements.} We are grateful to Sizheng Ma and
Zhen Pan for helpful discussions and valuable comments. Guangzhou Guo
acknowledges support from the China Postdoctoral Science Foundation (Grant
No.~2025M783414). This work was also supported in part by the NSFC (Grant
Nos.~12105191, 12275183, 12275184, 11875196).

\bibliographystyle{unsrturl}
\bibliography{ref}


\clearpage
\newpage\onecolumngrid

\begin{center}

\textit{{\Large Supplementary Material}}
\end{center}

\onecolumngrid
\setcounter{equation}{0} \setcounter{figure}{0} \setcounter{section}{0}
\setcounter{table}{0} \setcounter{page}{1} \makeatletter
\renewcommand{\theequation}{S\arabic{equation}}
\renewcommand{\thefigure}{S\arabic{figure}}
\renewcommand{\thetable}{S\arabic{table}} \vspace{-0.5cm}

\section*{Static Q-ball hairy BH solution}

The static solution used throughout the paper is obtained from the action
\eqref{eq:action-main} together with the spherically symmetric ansatz
\eqref{eq:bg-main}. Varying the action yields the Einstein, Maxwell, and
scalar equations,
\begin{align}
G_{ab}  &  =8\pi T_{ab},\nonumber\\
\nabla_{b}F^{ab}  &  =4\pi j^{a},\\
D_{a}D^{a}\Psi &  =\frac{1}{2}\,\frac{\partial U}{\partial|\Psi|^{2}}%
\,\Psi,\nonumber
\end{align}
with
\begin{equation}
T_{ab}=\frac{1}{8\pi}\left[  2F_{ac}F_{b}{}^{c}-\frac{1}{2}g_{ab}F^{2}+\left(
D_{a}\Psi\right)  ^{\ast}D_{b}\Psi+D_{a}\Psi\left(  D_{b}\Psi\right)  ^{\ast
}-g_{ab}\left(  |D\Psi|^{2}+\frac{1}{2}U\right)  \right]  ,
\end{equation}
and
\begin{equation}
j^{a}=-\frac{iq}{8\pi}\left[  \Psi^{\ast}D^{a}\Psi-\Psi\left(  D^{a}%
\Psi\right)  ^{\ast}\right]  .
\end{equation}
Substituting Eq.~\eqref{eq:bg-main} into these field equations reduces the
system to a set of coupled radial ordinary differential equations for
$m_{t}(r)$, $m_{r}(r)$, $V(r)$, and $\Psi_{0}(r)$. For the discussion of
boundary conditions, it is convenient to introduce the metric functions $N(r)$
and $\delta(r)$ through
\begin{equation}
e^{-2\delta(r)}N(r)=1-\frac{2m_{t}(r)}{r},\qquad N(r)=1-\frac{2m_{r}(r)}{r}.
\end{equation}
The explicit form of the resulting equations of motion can be found in
Ref.~\cite{Herdeiro:2020xmb}.

The static hairy BH solution is determined by imposing regularity at the event
horizon and asymptotic flatness at spatial infinity. At $r=r_{h}$, regularity
requires
\begin{equation}
N(r_{h})=0,\qquad V(r_{h})=0,\qquad\delta(r_{h})<\infty,\qquad\Psi_{0}%
(r_{h})<\infty.
\end{equation}
The condition $V(r_{h})=0$ fixes the regular horizon gauge. Together with the
residual $U(1)$ gauge freedom, this allows the static scalar profile $\Psi
_{0}(r)$ to be chosen real, as assumed in Eq.~\eqref{eq:bg-main}. One then
expands the fields in a regular Taylor series around $r=r_{h}$ and uses the
horizon data as shooting parameters.

At large radius, asymptotic flatness and scalar localization require
\begin{equation}
\delta(r)\to0,\qquad N(r)\to1-\frac{2M}{r}+\frac{Q^{2}}{r^{2}},\qquad
V(r)\to\Phi+\frac{Q}{r},\qquad\Psi_{0}(r)\to0,
\end{equation}
with $\Psi_{0}(r)$ decaying exponentially. The static solution is then
constructed by integrating outward from the horizon and tuning the shooting
parameters until these asymptotic conditions are satisfied.

The particular solution used in the main text has $(q,m,\beta,\gamma
)=(1.25,1.64,-301.88,9428.84)$ and $(Q,M)=(0.97,1)$, and belongs to the
dynamically unstable sector of the Q-ball hairy BH family. This static
solution is used as the initial equilibrium state in the nonlinear time-domain
evolution and as the background solution for the frequency-domain perturbation analysis.

\section*{Nonlinear time-domain evolution}

\label{app:evolution}

The static Q-ball hairy BH solution described above is used as the initial
equilibrium state for the full nonlinear evolution. The instability is seeded
solely by the numerical noise of the discretized system, with no external
perturbation imposed by hand.

Although the evolutions discussed in the main text are performed in spherical
symmetry, it is convenient to present the evolution system in standard $3+1$
form. The spacetime is decomposed as
\begin{equation}
\mathrm{d}s^{2}=-\alpha^{2}\mathrm{d}t^{2}+\gamma_{ij}(\mathrm{d}x^{i}%
+\beta^{i}\mathrm{d}t)(\mathrm{d}x^{j}+\beta^{j}\mathrm{d}t),
\end{equation}
where $\alpha$, $\beta^{i}$, and $\gamma_{ij}$ are the lapse, shift, and
spatial metric, respectively. The metric variables are evolved with the CCZ3
method. The Maxwell and charged scalar sectors is formulated in a $3+1$
decomposition, with evolution equations that are first order in time,
following the same evolution strategy as in
Refs.~\cite{Garcia-Saenz:2024beb,Garcia-Saenz:2025dsr,Garcia-Saenz:2025rbc}.

For the scalar sector, we introduce the complex momentum
\begin{equation}
\Pi=\frac{1}{\alpha}\left(  \partial_{t}-\mathcal{L}_{\beta}\right)  \Psi.
\end{equation}
Here $\nabla_{i}$ denotes the spatial covariant derivative compatible with
$\gamma_{ij}$. For the Maxwell sector, we use the standard $3+1$ projections
\begin{equation}
\mathcal{A}_{i}=\gamma_{i}{}^{a}A_{a},\qquad\mathcal{A}_{\phi}=-n^{a}%
A_{a},\qquad E_{i}=\gamma_{i}{}^{a}n^{b}F_{ab},\qquad B_{i}=\gamma_{i}{}%
^{a}n^{b}\tilde{F}_{ab},
\end{equation}
where $n^{a}$ is the future-directed unit normal to the spatial slice, and
$\tilde{F}_{ab}$ is the Hodge dual of $F_{ab}$. In practice, $B^{i}$ is
obtained from the spatial curl of the vector potential,
\begin{equation}
B^{i}=\epsilon^{ijk}\nabla_{j}\mathcal{A}_{k},
\end{equation}
with $\epsilon^{ijk}$ the spatial Levi-Civita tensor. Spatial indices are
raised and lowered with $\gamma_{ij}$.

The scalar field is then evolved according to
\begin{align}
(\partial_{t}-\mathcal{L}_{\beta})\Psi &  =\alpha\Pi,\nonumber\\
(\partial_{t}-\mathcal{L}_{\beta})\Pi &  =\nabla_{i}\!\left(  \alpha\nabla
^{i}\Psi\right)  +\alpha K\Pi-2iq\alpha\!\left(  \mathcal{A}_{i}\nabla^{i}%
\Psi+\mathcal{A}_{\phi}\Pi\right)  -\alpha\!\left[  q^{2}\!\left(
\mathcal{A}_{i}\mathcal{A}^{i}-\mathcal{A}_{\phi}^{2}\right)  \Psi+\frac{1}%
{2}\,\frac{\partial U}{\partial|\Psi|^{2}}\,\Psi\right]  .
\end{align}
where $K$ is the trace of the extrinsic curvature. The Maxwell sector is
evolved simultaneously through
\begin{equation}
\begin{aligned} (\partial_{t}-\mathcal{L}_{\beta})\mathcal{A}_{\phi} &= -\mathcal{A}^{i}\nabla_{i}\alpha +\alpha\left(K\mathcal{A}_{\phi}-\nabla_{i}\mathcal{A}^{i}\right),\\ (\partial_{t}-\mathcal{L}_{\beta})\mathcal{A}_{i} &= -\alpha\left(E_{i}+\nabla_{i}\mathcal{A}_{\phi}\right) -\mathcal{A}_{\phi}\nabla_{i}\alpha,\\ (\partial_{t}-\mathcal{L}_{\beta})E^{i} &= \alpha K E^{i} +\alpha\epsilon^{ijk}\nabla_{j}B_{k} -\epsilon^{ijk}B_{j}\nabla_{k}\alpha -4\pi\alpha j_{e}^{\,i}, \end{aligned}
\end{equation}
together with the electromagnetic Gauss-law constraint
\begin{equation}
\nabla_{i}E^{i}=4\pi\rho_{e}. \label{eq:gauss-constraint}%
\end{equation}
Here $\rho_{e}=-n_{a}j^{a}$ and $j_{e}^{\,i}=\gamma^{i}{}_{a}j^{a}$ are the
charge density and spatial current defined from the matter current $j^{a}$.

The matter source terms entering the CCZ3 equations are the standard $3+1$
projections of the stress tensor,
\begin{equation}
\rho=n^{a}n^{b}T_{ab},\qquad J_{i}=-\gamma_{i}{}^{a}n^{b}T_{ab},\qquad
S_{ij}=\gamma_{i}{}^{a}\gamma_{j}{}^{b}T_{ab},\qquad S=\gamma^{ij}S_{ij}.
\end{equation}
These quantities provide the matter source terms for the evolution of the
metric variables. In the simulations discussed here, spherical symmetry is
imposed, so the system reduces to a $1+1$-dimensional problem.

\section*{QNM filter}

\label{app:filter}

Our filtering procedure follows the rational QNM filter introduced in
Ref.~\cite{Ma:2022wpv}. For a given horizon time series $X(t)$, taken here to
be $\partial_{t}\psi^{R,I}(t,r_{h})$, we define its Fourier transform as
\begin{equation}
\tilde{X}(\omega)=\int_{-\infty}^{\infty}X(t)e^{i\omega t}\,dt.
\end{equation}
For a QNM with frequency $\omega_{n}$, the corresponding rational filter is
\begin{equation}
F_{n}(\omega)=\frac{\omega-\omega_{n}}{\omega-\omega_{n}^{\ast}}.
\end{equation}
For real $\omega$, this filter satisfies $|F_{n}(\omega)|=1$ and therefore
does not amplify high-frequency components. To remove a set $\mathcal{S}$ of
QNM contributions, we use the product filter
\begin{equation}
F_{\mathcal{S}}(\omega)=\prod_{n\in\mathcal{S}}F_{n}(\omega)=\prod
_{n\in\mathcal{S}}\frac{\omega-\omega_{n}}{\omega-\omega_{n}^{\ast}},
\end{equation}
and define the filtered signal by
\begin{equation}
X_{\mathcal{S}^{c}}(t)=\frac{1}{2\pi}\int_{-\infty}^{\infty}F_{\mathcal{S}%
}(\omega)\tilde{X}(\omega)e^{-i\omega t}\,d\omega.
\end{equation}

In the present case, the unstable spectrum satisfies $\omega_{n}=n\omega_{1}$.
Starting from the horizon signals $\partial_{t}\psi^{R,I}(t,r_{h})$, we obtain
$(\partial_{t}\psi^{R,I})_{1}$ by filtering out the components at $\omega
_{2},\ldots,\omega_{7}$, thereby isolating the $\omega_{1}$ contribution.
Likewise, $(\partial_{t}\psi^{R,I})_{2}$ are obtained by filtering out
$\omega_{1},\omega_{3},\ldots,\omega_{7}$, thereby isolating the $\omega_{2}$
contribution. Equivalently,
\begin{align}
(\partial_{t}\psi^{R,I})_{1}  &  :\quad\mathcal{S}=\{2,3,4,5,6,7\},\qquad
\nonumber\\
(\partial_{t}\psi^{R,I})_{2}  &  :\quad\mathcal{S}=\{1,3,4,5,6,7\}.
\end{align}

In the present setting, the filter is used only as a diagnostic of the mode
content within the unstable growth window. Over a finite time window, the
filtered signal can be affected by boundary-induced artifacts, and the
rational filter can also modify the timing and phase structure of the signal.
As a result, the filtered signals $(\partial_{t}\psi^{R,I})_{1}$ and
$(\partial_{t}\psi^{R,I})_{2}$ are useful for identifying the presence of the
$\omega_{1}$ and $\omega_{2}$ contributions and for indicating which of them
is more important in a given component, but they are not suitable for
extracting a precise ratio between the linear and quadratic contributions. For
that purpose, we instead rely on multimode fits to the horizon signals.

\section*{Frequency-domain calculation}

\label{app:fd}

Since the nonlinear evolutions discussed in the main text are performed
in spherical symmetry, we restrict the perturbation analysis to purely
radial perturbations of the unstable Q-ball hairy BH. To adopt the same gauge choice as in the time-domain evolution, we impose the
Lorentz gauge throughout the perturbative analysis,
\begin{equation}
\nabla_{a}A^{a}=0.
\end{equation}
As a consequence, although the radial component of the vector potential
vanishes in the static solution, it must be retained at the perturbative
level. We expand the
fields $X=\{m_{t},m_{r},V,A_{r},\Psi,\Psi^{\ast}\}$ order by order
in a bookkeeping parameter $\epsilon$ as
\begin{equation}
X(t,r)=X_{0}(r)+\epsilon X_{1}(t,r)+\epsilon^{2}X_{2}(t,r)+\cdots,
\end{equation}
where $X_{0}(r)$ denotes the static solution, while $X_{1}(t,r)$
and $X_{2}(t,r)$ are the first- and second-order perturbations, respectively.
For convenience, we introduce
\begin{equation}
\Psi^{+}\equiv\Psi=\Psi^{R}+i\Psi^{I},\qquad\Psi^{-}\equiv\Psi^{\ast}=\Psi^{R}-i\Psi^{I}.
\end{equation}

To compute the linear and quadratic QNMs, we separate the time dependence
according to
\begin{equation}
X_{1}(t,r)=e^{-i\omega_{1}t}X_{1}(r),\qquad X_{2}(t,r)=e^{-i\omega_{2}t}X_{2}(r),
\end{equation}
with $\omega_{2}=2\omega_{1}$ for the self-coupled quadratic channel.
Substituting these ansätze into the full coupled perturbation equations
yields, at first order, the homogeneous eigenvalue problem
\begin{equation}
\mathcal{L}(\omega_{1})X_{1}(r)=0,\label{eq:linear}
\end{equation}
and, at second order, the sourced equations
\begin{equation}
\mathcal{L}(\omega_{2})X_{2}(r)=\mathcal{S}_{2}[X_{1}(r),X_{1}(r)].\label{eq:quadratic}
\end{equation}
Here $\mathcal{L}(\omega)$ is the radial differential operator obtained
from the linearized field equations, while $\mathcal{S}_{2}$ is the
quadratic source constructed from the first-order solution. The first-order
problem determines the unstable frequency $\omega_{1}$ and eigenfunction
$X_{1}(r)$. The second-order problem then uses $X_{1}(r)$ as a known
source and yields the quadratic mode $X_{2}(r)$ at frequency $\omega_{2}=2\omega_{1}$.

At linear order, the radial perturbation equations can be reduced
to four coupled equations for $\{V_{1}(r),A_{r1}(r),\Psi_{1}^{+}(r),\Psi_{1}^{-}(r)\}$.
Near the event horizon, regularity in ingoing Eddington-Finkelstein
coordinates requires
\begin{equation}
\Psi_{1}^{\pm}(r)\sim e^{-i\omega r_{\ast}},\qquad r\rightarrow r_{h},
\end{equation}
where $r_{\ast}$ is the tortoise coordinate, while the perturbation
of the vector potential must satisfy
\begin{equation}
V_{1}(r)\sim e^{-i\omega r_{\ast}}, \qquad
A_{r1}(r)\sim \frac{e^{-i\omega r_{\ast}}}{r-r_{h}},
\qquad r\to r_{h}.
\end{equation}
At spatial infinity, the three equations decouple asymptotically,
and the solutions behave as
\begin{equation}
\Psi_{1}^{\pm}(r)\sim B_{\pm}e^{-k_{\pm}r}+C_{\pm}e^{k_{\pm}r},\qquad V_{1}(r)\sim a_{0}+\mathcal{O}(r^{-1}),\qquad A_{r1}(r)\sim\frac{b_{1}}{r}+\mathcal{O}(r^{-2}),\qquad r\rightarrow\infty,
\end{equation}
with
\begin{equation}
k_{\pm}^{2}=\frac{m^{2}}{2}-\left(\omega\pm q\Phi\right)^{2}.
\end{equation}
For the unstable mode relevant here, we impose the decaying branch
at spatial infinity, which corresponds to setting $C_{\pm}=0$ \cite{Chen:2025oxo}.

In the original Schwarzschild-like radial coordinate, the unstable
mode eigenfunctions exhibit exponential behavior near the event horizon,
which tends to reduce numerical stability. To factor out this horizon
behavior, we adopt hyperboloidal coordinates $(\tau,x)$ \cite{Zenginoglu:2007jw,Zenginoglu:2011jz,PanossoMacedo:2023qzp},
defined by
\begin{equation}
t=\tau-H(x),\qquad r=\frac{2r_{h}}{1-x},
\end{equation}
which map the exterior region $r\in\lbrack r_{h},\infty)$ onto the
finite interval $x\in\lbrack-1,1]$. In the present calculation, the
height function $H(x)$ is chosen to factor out the asymptotic behavior
at the event horizon. In particular, we take
\begin{equation}
H(x)=\frac{r_{h}e^{\delta(-1)}}{2N^{\prime}(-1)}\log(x+1),
\end{equation}
where $N(x)\equiv N[r(x)]$ and $\delta(x)\equiv\delta\lbrack r(x)]$.
We then define the rescaled first-order fields
\begin{equation}
\tilde{X}_{1}(x)=e^{i\omega_{1}H(x)}X_{1}(r),
\end{equation}
for which the horizon behavior has been factored out. The first-order
equation \eqref{eq:linear} is thereby transformed into
\begin{equation}
\tilde{\mathcal{L}}(\omega_{1})\tilde{X}_{1}(x)=0.\label{eq:tlinear}
\end{equation}

At second order, we similarly define
\begin{equation}
\tilde{X}_{2}(x)=e^{i\omega_{2}H(x)}X_{2}(r).
\end{equation}
As at first order, the six second-order perturbation equations can
be reduced to four independent equations for $\{\tilde{V}_{2}(x),\tilde{A}_{r2}(x),\tilde{\Psi}_{2}^{R}(x),\tilde{\Psi}_{2}^{I}(x)\}$.
Equation~\eqref{eq:quadratic} then becomes
\begin{equation}
\tilde{\mathcal{L}}(\omega_{2})\tilde{X}_{2}(x)=\tilde{\mathcal{S}}_{2}[\tilde{X}_{1}(x),\tilde{X}_{1}(x)],\label{eq:tquadratic}
\end{equation}
where the source term $\tilde{\mathcal{S}}_{2}$ is regular at both
boundaries.

To solve Eqs.~\eqref{eq:tlinear} and \eqref{eq:tquadratic}, we
employ a Chebyshev spectral method. We collect the unknown first-
and second-order fields into
\begin{equation}
\mathcal{F}=\{\tilde{V}_{1},\tilde{A}_{r1},\tilde{\Psi}_{1}^{R},\tilde{\Psi}_{1}^{I},\tilde{V}_{2},\tilde{A}_{r2},\tilde{\Psi}_{2}^{R},\tilde{\Psi}_{2}^{I}\},
\end{equation}
and approximate each component by a truncated Chebyshev expansion,
\begin{equation}
\mathcal{F}^{(k)}(x)\approx\sum_{i=0}^{N_{x}-1}a_{i}^{(k)}T_{i}(x),
\end{equation}
where $T_{i}(x)$ are Chebyshev polynomials, $a_{i}^{(k)}$ are the
spectral coefficients, and $N_{x}$ sets the spectral resolution.
We then discretize Eqs.~\eqref{eq:tlinear} and \eqref{eq:tquadratic}
on the collocation grid in the compact coordinate $x$.

For the first-order problem, the discretization of Eq.~\eqref{eq:tlinear}
yields a nonlinear algebraic system for the spectral coefficients
together with the eigenfrequency $\omega_{1}$. To remove the overall
scaling freedom of the homogeneous eigenvalue problem, we impose the
normalization condition
\begin{equation}
\tilde{\Psi}_{1}^{R}(-1)=1.
\end{equation}
The resulting nonlinear system is solved in the complex plane by means of the
Newton-Raphson method. At each iteration, the residual equations are
linearized around the current estimate, and the corresponding linear system is
solved to obtain the correction. The iteration is terminated once the
difference between successive iterates falls below $10^{-10}$.
Once the first-order solution has been obtained, the second-order
equation \eqref{eq:tquadratic} becomes a linear algebraic system
for the second-order spectral coefficients, since the source term
is then known and $\omega_{2}=2\omega_{1}$ is fixed. Solving this
system yields the quadratic mode profile.

Finally, Eq.~\eqref{eq:pert-main} implies
\begin{equation}
\psi_{i}^{R}=\Psi_{i}^{R},\qquad\psi_{i}^{I}=\Psi_{i}^{I},\qquad i=1,2.
\end{equation}
Therefore, the frequency-domain counterpart of the ratio introduced
in the main text is
\begin{equation}
\mathcal{R}=\frac{\tilde{\Psi}_{2}^{R}(x)}{\left[\tilde{\Psi}_{1}^{I}(x)\right]^{2}},
\end{equation}
evaluated at the horizon. Because $\mathcal{R}$ is invariant under
the rescalings $X_{1}\rightarrow aX_{1}$ and $X_{2}\rightarrow a^{2}X_{2}$,
it does not depend on the normalization chosen for the first-order
eigenfunction.

\end{document}